\documentclass{article}

\newcommand{\beq}{\begin{equation}}
\newcommand{\eeq}{\end{equation}}
\newcommand{\zpr}{\mbox{$Z'$}}
\newcommand{\mzp}{\mbox{$M_{Z'}$}}
\newcommand{\nur}{\mbox{$\nu_R$}}
\newcommand{\upr}{\mbox{$U(1)'$}}

\def\mxth{\mathsurround=0pt }
\def\xversim#1#2{\lower2.pt\vbox{\baselineskip0pt \lineskip-.5pt
  \ialign{$\mxth#1\hfil##\hfil$\crcr#2\crcr\sim\crcr}}}

\newcommand{\neff}{\mbox{$\Delta N_\nu$}}

\newcommand{\bd}{\begin{equation}}
\newcommand{\ed}{\end{equation}}

\newcommand{\bea}{\begin{eqnarray}}
\newcommand{\eea}{\end{eqnarray}}
\newcommand{\ba}{\begin{array}}
\newcommand{\ea}{\end{array}}

\begin{document}


%
%

\title{\normalsize ALTERNATIVES TO THE SEESAW: EXTRA $Z$'S AND CONSTRAINTS ON LARGE
EXTRA DIMENSIONS\footnote{Presented at {\em Neutrinos and Implications for Physics
Beyond the Standard Model}, Stony Brook, October 2002.} }

%
\author{\normalsize PAUL LANGACKER \\
Department of Physics, University of Pennsylvania \\
Philadelphia, PA 19104, USA\\
UPR-1035T}

\date{\today}
\maketitle


\begin{abstract}
Alternatives to the traditional grand unified theory seesaw for neutrino masses
are briefly described. These include the possibility of large extra dimensions
and various possibilities for models involving an extra \upr  \  gauge symmetry.
The difficulty of observing Majorana phases in neutrinoless double beta decay
is also briefly commented on.
\end{abstract}

\section{Introduction}	
\begin{itemize}
  \item The GUT seesaw
\item {Large Extra Dimensions}
\item {Extra \upr s}
\item {Implications of \upr \ for Neutrino Mass}
\item {Primordial Nucleosynthesis}
\item {Natural Decoupling of the \nur}
\item {A Comment on Majorana Phases in Neutrinoless Double Beta Decay}
\end{itemize}

\roman{footnote}
\section{The GUT Seesaw}

The grand unified theory seesaw model~\cite{seesaw} is an elegant
mechanism for generating small Majorana neutrino masses, which leads
fairly easily to masses in the correct range.  It  also provides a
simple framework for leptogenesis~\cite{leptogenesis}, in which the
decays of heavy Majorana neutrinos produce a lepton asymmetry, which is
later partially converted to a baryon asymmetry by electroweak sphaleron
effects.

However, the expectation of the simplest grand unified theories is
that the quark and lepton mixings should be comparable and that the
neutrino mixings (i.e., the mismatch between the neutrino and charged
lepton mixings) should be small, rather than the large mixings that
are observed. This can be evaded in more complicated GUTs, e.g.,
those involving ``lopsided'' mass matrices~\cite{lopsided}, in which
there are large mixings in the right-handed charge $-1/3$ quarks
(where it is unobservable) and in the left-handed charged leptons;
those with complicated textures~\cite{textures} for the heavy
Majorana neutrino mass matrix; or in Type II seesaw models involving
a Higgs triplet~\cite{typeII},
but the need to do so makes the GUT
seesaw concept less compelling. Furthermore, a number of promising
extensions of the standard model or MSSM do not allow the canonical
GUT seesaw. For example, the large Majorana masses needed are often
forbidden, e.g., by extra \upr \ symmetries predicted in many string
constructions and some GUTs. Similarly, it is difficult to accomodate
traditional grand unification (especially the needed adjoint and high
dimension Higgs multiplets needed for GUT breaking and the seesaw)
in simple heterotic string constructions. Such constructions also
tend to forbid direct Majorana mass terms and large scales.  Finally,
the active-sterile neutrino mixing~\cite{sterile}
required in the
four-neutrino schemes~\footnote{It is difficult to accomodate the data
in such schemes. See, for example,  ref~\cite{fournu}.} motivated by
the LSND experiment~\cite{LSND} is difficult to implement in canonical
seesaw schemes.

For all of these reasons it is useful to explore alternatives to the
canonical GUT seesaw with small Dirac and/or Majorana masses.

In this talk I will briefly describe some possible alternatives. These
include large extra dimensions~\cite{led1}, which can lead to small
(volume-suppressed) Dirac masses~\cite{LED,DLP}; models involving
an extra \upr \ gauge symmetry broken at the TeV scale, for which
small Majorana masses can be generated by an extended ``TeV'' seesaw
mechanism~\cite{TEVseesaw}; models in which small masses (Dirac,
Majorana, or both) are generated by higher dimensional operators,
such as can occur if an extra \upr \ is broken at an intermediate
scale~\cite{HDO}; and the implications of a $\upr \times \upr$ gauge
symmetry, which can lead to the decoupling of a TeV-scale \zpr \
from right-handed neutrinos $\nu_R$\footnote{Other directions include
loop-induced masses~\cite{loop} or masses associated with $R$-parity
breaking in supersymmetry~\cite{numass}.}. I also mention a new
detailed calculation of big bang nucleosynthesis constraints on \upr \
models involving Dirac neutrinos~\cite{BLL}, which motivate the need
for $\nu_R$\ decoupling, and comment on the possibility of observing
Majorana phases in neutrinoless double beta decay~\cite{phase}.

\section{Large Extra Dimensions}
The possibility of extra space dimensions is suggested by
string/M theory, which implies 5 or 6 new dimensions,
and by bottom-up motivated brane world scenarios. These could all be
very small, with radii $R$ of order the inverse Planck scale
$R \sim 1/\bar{M}_{Pl}$, where $\bar{M}_{Pl}= 1/\sqrt{8 \pi G_N} 
\sim 2.4 \times 10^{18}$
GeV. However, it is possible that at least some of them are larger~\cite{led1}.
A major motivation is the hierarchy problem, i.e., why is $\bar{M}_{Pl}$ so
much larger than the electroweak scale? One possibility is that the true
fundamental scale of nature, $M_F$, is much smaller than $\bar{M}_{Pl}$,
e.g., $M_F \sim 1-100$ TeV, and that there are $\delta$  roughly comparable
extra dimensions
with a volume $V_\delta \sim R^\delta$ that is very large compared
to $M_F^{-\delta}$ (note that this introduces a new hierarchy problem
to explain why $R$ is so large). Gravity is modified on scales
smaller than $R$, so that
\beq  \bar{M}_{Pl}^2 = M_F^{2+\delta} V_\delta \gg M_F^2. \eeq
For $M_F = 100$ TeV, this implies $R \sim 10^8$ cm for $\delta = 1$, which
is clearly excluded, or $R \sim 5 \times 10^{-6}$ cm for $\delta = 2$.
These ideas could be implemented in
 a brane scenario in which the gravitons propagate in
the extra-dimensional bulk, and the other matter fields are confined to 
four-dimensional branes. The implications
for the  production or exchange of gravitons and their Kaluza-Klein excitations
(of mass $k/R,\  k=1, \cdots )$ at colliders as well as in cosmology and
supernovae have been studied extensively~\cite{gravitons}.
 Cavendish-type
laboratory experiments imply that the largest extra dimension is smaller than
around 0.02 cm~\cite{cavendish}, with a future sensitivity to around 0.005 cm.

A number of authors~\cite{LED} have discussed the implications of large extra dimensions
for neutrino masses in a scenario in which one extra dimension is much
larger than the other $\delta - 1$ (but still allowing the other dimensions
to be much larger than $M_F^{-1}$). The idea is that sterile ($SU(2)$-singlet)
neutrinos $N_{L,R}$ could propagate in the bulk along with gravitons, with 
other matter confined to the brane\footnote{This does {\em not} occur in 
most specific string constructions, however. Rather, the $N_R$ are generally
confined to branes.}. One of the $N_R$ becomes the right-handed partner
of an active ($SU(2)$-doublet) neutrino on the brane. The advantage is that the
Dirac mass coupling the active neutrino to its partner is suppressed by a
volume factor, so that it is naturally small, i.e.,
\beq
m_D \sim h v M_F/\bar{M}_{Pl}, \eeq
where $h$ is a Yukawa coupling and $v$ is the electroweak scale. For
$h \sim 1$ and $M_F \sim$ 100 TeV this yields $m_D$
around $10^{-2}$ eV. 
Unfortunately,  no light is shed on the
family structure, mass hierarachy, or mixings.
In addition to predicting a small (or even too small) Dirac mass,
such schemes predict the existence of Kaluza-Klein excitations of the sterile
$N_{L,R}$, with masses around $k/R$, $k=1, \cdots $. For $R^{-1} > $ O(1~eV), for
example,  $R < $ O($10^{-5}$ cm). These excitations are sterile, 
leading to the possibility  of oscillations of active neutrinos
into these sterile states. Unlike most other schemes for active-sterile
oscillations~\cite{sterile}, however, there is a conserved lepton number,
at least in the simplest versions, so that there would be no 
neutrinoless double beta decay\footnote{Active-sterile oscillations require
the existence of two types of mass terms of comparable magnitude: the 
Higgs-generated Dirac masses coupling active and sterile states, and mass terms
connecting sterile neutrinos. In most other models, the latter are given by
 Majorana masses, but in the present case they are the lepton number-conserving
Kaluza Klein masses.}.

Some of the early papers attempted to interpret the atmospheric and/or Solar
neutrino oscillations as active into sterile. This is no longer
viable in view of Superkamiokande, MACRO, and SNO data. Also,
some authors have combined the notion of small Dirac masses from large dimensions
with Majorana mass terms generated by some other mechanism. In~\cite{DLP}
we analyzed a minimal viable scheme in which there are three active neutrinos on
the brane and three sterile neutrinos, with their towers of Kaluza-Klein
excitations, in the bulk, and in which the only mass terms in the effective
four-dimensional
theory were the lepton number conserving but naturally small Dirac masses and
the Kaluza-Klein masses. The Solar and atmospheric neutrino oscillations were
assumed to be primarily due to the dominant active-active oscillations. The
only roles of the extra dimensions were (a) to provide a natural explanation
for small Dirac masses (but not to explain the details of the mass hierarchy
and mixings), and (b) to allow subdominant active to Kaluza-Klein sterile
transitions as a  perturbation. Results from reactor, accelerator, Solar,
and atmospheric experiments were used to constrain the amount of
active-sterile mixing, and therefore to bound the radius $R$ of the largest
dimension in these schemes. The results
 in Table \ref{tab:rabounds}, taken from~\cite{DLP}, assuming
hierarchical, inverted, and degenerate mass patterns, are quite
stringent. For the hierarchical scheme, $R > 8.2 \times 10^{-5}$ cm and
$1/R < 0.24$ eV, with stronger constraints for the other mass  patterns.
These limits are stronger than those obtainable in Cavendish
type experiments (which, however, are still relevant in  schemes not involving
sterile neutrinos in the bulk), and
weaker than but complementary to constraints from cosmology~\cite{cosmology} and
supernova energy loss~\cite{SNbound}. 

One of the motivations of the study was
to generalize the marginally-successful $3+1$ schemes~\cite{fournu}
 proposed to
accomodate the LSND results, with multiple sterile states. 
However, despite the
large number of sterile Kaluza-Klein excitations, this simplest scheme cannot accomodate
the LSND results because of cancellations between the effects of mixings with
the sterile states in the three Kaluza-Klein towers. More
complicated schemes are still possible~\cite{DLP}.

\begin{table}[h]
\caption{Upper bounds on $R$ (cm) at 90\% c.l.
and the corresponding lower bounds on $1/R$ (eV) from various
measurements. }
{\begin{tabular}{cccc}
\hline
\multicolumn{4}{c}{Experimental Bounds} \\
\hline
Experiment  & Hierarchical & Inverted & Degenerate \\
 & (cm, eV)& (cm, eV)& (cm, eV) \\
\hline
CHOOZ & ($9.9 \times 10^{-4}, 0.02$) & ($3.3\times 10^{-5}, 0.60$) & ($1.8\times 10^{-6},10.9 $)\\
BUGEY & none & ( $4.3\times 10^{-5},0.46$)& ($2.4\times 10^{-6},8.3 $) \\
CDHS & none &  none &  ($5 \times 10^{-6},4$) \\
Atmospheric & ($8.2 \times 10^{-5}, 0.24$) & ($6.2\times 10^{-5}, 0.32$) & ($4.8\times 10^{-6}, 4.1$)\\
Solar & ($1.0 \times 10^{-3}, 0.02$) & ($8.9\times 10^{-5}, 0.22$) & ($4.9\times 10^{-6}, 4.1$)\\
\hline
\hline
\end{tabular}}
\label{tab:rabounds}
\end{table}

\section{Extra \upr s}
Additional heavy \zpr \ gauge bosons~\cite{ell} are predicted in many
superstring~\cite{string} and grand unified~\cite{review} theories, and
also in models of dynamical symmetry breaking~\cite{DSB}.
If present at a scale of a TeV or so they could provide
a solution\footnote{This has the advantage of not
introducing the undesirable domain walls generally found in 
the NMSSM~\cite{NMSSM},
 in which the effective $\mu$ term is generated by the expectation value
of a new standard model singlet but there is no \upr \ gauge symmetry.}
 to the $\mu$ problem~\cite{muprob} and other
problems of the minimal supersymmetric standard model (MSSM)~\cite{general}.
This can be implemented in both supergravity~\cite{muprob} and gauge-mediated
models of supersymmetry breaking~\cite{gaugeupr}.
Current limits from collider~\cite{explim,LEPmixing} and precision~\cite{indirect}
experiments are model dependent, but generally imply that $M_{Z'} > (500-800)$ GeV
and that the $Z-Z'$ mixing angle is smaller than a few $\times 10^{-3}$.
A \zpr could be relevant to the NuTeV experiment~\cite{NuTeV} and,
if the couplings are not family universal~\cite{nonuniversal,indirect}, to the
anomalous value of the forward-backward asymmetry $A^b_{FB}$~\cite{afb}.
Earlier hints of a discrepancy in atomic parity violation~\cite{APV} have largely
disappeared due to improved calculations of radiative corrections~\cite{interp}.
A \zpr \ lighter than a TeV or so should be observable at Run II
at the Tevatron. Future colliders should be able to observe
a \zpr \ with mass up to around 5
TeV and perform diagnostics on the couplings up to a few
TeV~\cite{collider}.

The $E_6$ grand unified group contains two extra \upr \ factors, one or both
of which could survive to low energies. However,  canonical grand unification
is unlikely to lead to a TeV-scale \zpr \ because (a) there is no special reason for
the \zpr \ to be light without a second fine-tuning, (b) chiral colored scalars
with masses at the \zpr \ scale \ would mediate too rapid proton decay unless
the
$E_6$ Yukawa relations are broken. The latter could occur if the larger group
were broken by underlying string or higher-dimensional effects. In any case,
$E_6$ \upr \ models are generally viewed simply as examples of 
anomaly-free \upr \ gauge couplings for
the ordinary particles and for the new exotic particles that are typically
present in \upr \ theories, and not as full grand unified theories~\cite{LW}.

Detailed quasi-realistic heterotic~\cite{CCEEL} and open-string~\cite{CSU,CLS}
constructions (including the analysis of the
relevant flat directions~\cite{flatness})
usually lead to additional \upr \ gauge
factors and associated exotic chiral supermultiplets. These typically survive
to the TeV scale~\cite{string} \footnote{The natural expectation is $\mzp \sim
M_Z$. A TeV scale \zpr can occur by a moderate tuning of
parameters~\cite{muprob,LW,CCEEL}, or more naturally
if the breaking would occur along a  direction that becomes flat
in the limit in which
a small Yukawa coupling vanishes~\cite{secluded}.}, 
although some flat directions allow
breaking at an intermediate scale (e.g., $10^{12}$ GeV) if the spectrum of
the theory allows breaking along a $F$ and $D$-flat 
direction~\cite{CCEEL,intermediate}. The \zpr \ couplings are
typically family-nonuniversal, which could have implications
for $A^b_{FB}$~\cite{indirect}, and lead to associated
flavor changing
\zpr
\ and
 (including $Z-\zpr$ mixing) $Z$ couplings~\cite{nonuniversal}. 
The experimental
constraints from $K$ and $\mu$ decays imply that the couplings
of the first two families are most likely universal, but non-universality 
for the
third family could be of importance for $B$ and $\tau$ decays, as well as
leading to second order effects in $K$ decays,
$\mu \rightarrow 3e$, etc.

\section{Implications of \upr \ for Neutrino Mass}

Let \nur \ be a sterile ($SU(2)$-singlet) neutrino.
This would be the (``right-handed'') partner of an ordinary ($SU(2)$-doublet)
neutrino $\nu_L$ for a Dirac neutrino or the superheavy Majorana neutrino
in a seesaw model~\cite{seesaw}. In four-neutrino interpretations
of the LSND experiment, there must be significant mixing between
the $\nu_L$ and the left-handed CP-conjugate of the \nur, requiring
Dirac mass
terms to connect the doublet with the singlet and also  mass
terms\footnote{In
most models these are Majorana. However, it is also possible to introduce
lepton-number conserving singlet-singlet mass terms, as in the case of the
simple large extra dimension constructions.} to connect  two singlets (or
two doublets). These must both be small and comparable.
 
If the  \nur \ 
in a model  with an electroweak or TeV-scale \zpr \  carry a non-zero
\upr \ charge, then the \upr \ symmetry forbids them from obtaining
a Majorana mass much larger than the \upr-breaking scale. This
would forbid a conventional neutrino seesaw model~\cite{seesaw}.

In this case, it might still be possible to generate small Majorana
masses for the ordinary (active) neutrinos by some sort of
TeV-scale seesaw mechanism in which there are additional mass 
suppressions~\cite{TEVseesaw}, in schemes in which the masses are
suppressed by supersymmetry breaking~\cite{susylowmass}, or schemes involving a
superheavy Higgs triplet~\cite{triplet,typeII}. 
Another possibility is that
there are no Majorana mass terms, and that the neutrinos have
Dirac masses which are small for some reason, such as 
volume suppressions in theories with large
extra dimensions~\cite{LED,DLP} or other suppression
mechanisms~\cite{smallDirac,susylowmass}.

A small Dirac mass
could also arise
due to higher dimensional operators associated with an intermediate
scale~\cite{HDO}. In the \upr \ context these could occur if an extra \upr \ is
associated with a potential which is $F$ and $D$ flat at tree level,
with the flatness lifted by higher dimensional operators or
loop effects~\cite{intermediate}. (For the present application, a second
\upr \ could survive to low energies). For example, if there are
two standard model singlet fields $S_{1,2}$ whose expectation values
would break the \upr \ and which have opposite signs for their \upr \
charges\footnote{Whether such pairs actually occur in a given string
construction, for example, depends on the flat directions of the construction
and on the model~\cite{CCEEL,CSU,CLS}.}, then 
$F$-flatness would imply a tree level potential
\begin{equation}
V(S_1,S_2) = m_1^2 |S_1^2| + m_2^2 |S_2^2| + \frac{g'^2 Q'^2}{2}
(|S_1^2| - |S_2^2|)^2,
\label{potential}
\end{equation}
where $g'$ and $Q'$ are the \upr \ gauge coupling and charge (assumed
equal and opposite for simplicity), and $m_i^2$ are their soft mass-squares,
assumed to be of the order of the supersymmetry breaking (and electroweak)
scale $m_{soft}$. 
 For
 $m_1^2 < 0$ but
 $m_1^2 + m_2^2 > 0$, the minimum occurs for
 $\langle S_1 \rangle \sim m_{soft}$ and $\langle S_2 \rangle = 0$.
 However, for $m_1^2 + m_2^2 < 0$,
 the minimum occurs along the $D$-flat direction
 $\langle S_1 \rangle = \langle S_2 \rangle \equiv \langle S \rangle$,
 for which
 \begin{equation}
 V(S) = m^2 |S^2|,
 \label{dflat}
 \end{equation}
 where $m^2 = m_1^2 + m_2^2<0$. This potential appears to be unbounded from
below.
 However, the potential will be stabilized by loop corrections, which will
 cause the $m_i^2$ to become positive at a high enough scale, and/or by higher
 dimensional operators, so that $S$ obtains an expectation value
 at an intermediate scale, e.g., $10^{12}$ GeV.
Neutrino (and other fermion) masses could be generated by higher dimensional
operators in the superpotential. For example, the operators (suppressing 
family indices)
\begin{equation}
W \sim \hat{H}_2 \hat{L}_L \hat{\nu}^c_L
            \left( {\hat{S}\over {\cal M}} \right)^{P_D}
	    \label{ops}
	    \end{equation}
could lead to small Dirac neutrino masses.
In (\ref{ops}), the hats refer to superfields; $H_2$, $L_L$, $\nu_L^c$, and
$S$  are respectively the up-type Higgs doublet, the lepton doublet, the
left-handed CP conjugate of the $\nu_R$, and a singlet field; 
$\cal{M}$ is the Planck scale, and $P_{D}$ are nonnegative
integers. For $\langle S \rangle \ll \cal M$ and $P_D > 0$,
where $\langle S \rangle$ is the vacuum expectation value of the
scalar component of $\tilde{S}$, this implies a
suppressed Dirac  mass. Variants of the model, e.g., not involving a low
energy \upr \ or in which the \nur \ is not charged,
can also lead naturally to a scenario with significant
ordinary-sterile neutrino mixing~\cite{HDO}, as suggested by LSND.

\section{Primordial Nucleosynthesis}
In the purely Dirac case, there are six nearly massless two-component
neutrinos, the three active $\nu_L$ and their  partners $\nu_R$.
Such light Dirac neutrinos (i.e., with mass less than an eV or so) in the
standard model or MSSM are essentially sterile, except for the tiny
effects associated with their masses and Higgs couplings, which are much
too small to produce them in significant numbers prior to
nucleosynthesis or in a supernova.  However,  the superweak interactions
of these states due to their coupling to a heavy
$Z'$ (or a heavy $W'$ in the $SU(2)_L \times SU(2)_R \times U(1)$ extension of the
standard model) might be sufficient to create
them in large numbers in the early
universe~\cite{steigman,dolgov}.
The implications of such interactions were worked out qualitatively  by
Steigman, Olive,
and Schramm~\cite{steigman}. At high temperatures the \nur \ would
have been produced by the \zpr \ interactions. However, assuming
that the \zpr \ is heavier than the $W$ and $Z$ and that the $Z-Z'$ mixing
is small, the \nur \
decoupled earlier than ordinary neutrinos. As the temperature dropped
further, massive particles such as heavy quarks, pions, and muons subsequently
annihilated, and the light quarks and gluons were confined at the time
of the quark-hadron phase transition, reheating
the ordinary neutrinos and other particles in equilibrium, but not the \nur.
 Thus,  when the temperature dropped to a few MeV the amount of
extra relativistic energy density associated with the \nur \
would be somewhat suppressed.  Their contribution is usually
parametrized by the number \neff \ of additional 
active neutrinos 
that would yield the same contribution to the energy
density. Additional relativistic energy density would increase the expansion
rate, leading to a larger prediction for the primordial $^4He$ abundance.
The observed amount is still rather uncertain, but
typical estimates of the upper limit on \neff \ are in the range
$\neff < (0.3-1)$~\cite{dolgov}.

Recently, the first complete calculations of this effect were made,
for the example of the \upr  \ charges in $E_6$ models, for various
assumptions concerning the quark-hadron transition and the $Z-Z'$
mixing~\cite{BLL} \footnote{Earlier estimates, cited in~\cite{BLL}, did
not consistently include the effects of all of the particles present at
a given temperature on the \nur \ interaction rate.}.
The $E_6$ model actually involves two \upr \ factors,
and it was assumed that only one linear combination
of the charges survived to low energies. For most of the 
cases the \zpr \ mass and mixing constraints
from nucleosynthesis are in the multi-TeV range, and 
are much more stringent than the existing laboratory
limits from searches for direct production or from precision electroweak
data, and are comparable to the ranges that may ultimately be probed
at proposed colliders.
They are qualitatively similar to
the limits from energy emission from Supernova 1987A~\cite{supernova},
but somewhat more stringent for
$\neff<0.3$, and have entirely different theoretical and systematic uncertainties.

\section{Natural Decoupling of the \nur}
Both the nucleosynthesis and supernova constraints can be evaded
if the \nur \ does not couple to the \zpr. This can in fact
occur naturally in classes of models in which
one combination of  the two \upr \  charges is broken at a
large scale associated with an $F$ and $D$-flat 
direction~\cite{decouplingmodel},
leaving a light \zpr \ which decouples from the 
$\nu_R$\footnote{A large Majorana
mass for the $\nu_R$ may still be forbidden in the model.}.

To illustrate this, consider the $E_6$ couplings. The
two \upr \ factors are  referred to as
$U(1)_\chi$, associated with $SO(10) \rightarrow SU(5) \times U(1)_\chi$,
and $U(1)_\psi$, corresponding to $E_6 \rightarrow SO(10) \times U(1)_\psi$.
Each family contains two standard model singlet fields, \nur \ and
an $SO(10)$-singlet  $s_L$. One can also add to the model
pairs of $ \nur + \nu^*_R$ and  $s_L+ s^*_L$ 
from $27+27^*$-plets. Then, the \upr \ $D$ terms of the potential are
\bea
 V_\chi+V_\psi &=& \frac{g'^2}{2} \left[ 
\frac{5}{2\sqrt{10}} (|\tilde{\nu}_R|^2 - |\tilde{\nu}^*_R|^2) \right]^2
\nonumber \\
 & + &
\frac{g'^2}{2} \left[ 
 \frac{1}{\sqrt{24}} (-|\tilde{\nu}_R|^2 + |\tilde{\nu}^*_R|^2
-4|\tilde{s}_L|^2 + 4|\tilde{s}^*_L|^2) \right]^2,
\label{dterm}
\eea
where the tildes represent scalar fields, a sum over each type of
scalar is implied, and I
have assumed equal gauge couplings for simplicity.
The potential is clearly $D$-flat for 
$|\tilde{\nu}_R|^2 = |\tilde{\nu}^*_R|^2 \equiv |\tilde{\nu}|^2 $ and
$|\tilde{s}_L|^2 = |\tilde{s}^*_L|^2 \equiv |\tilde{s}|^2 $.
I will assume that the potential is also $F$-flat along the this direction.
The potential along the flat directions is then
\beq V(\tilde{\nu}, \tilde{s}) = m^2_{\tilde{\nu}}  |\tilde{\nu}^2|
+ m^2_{\tilde{s}} |\tilde{s}^2|, \eeq
in analogy with (\ref{dflat}), where $ m^2_{\tilde{\nu}}$ and $m^2_{\tilde{s}}$
are respectively the sum of the mass-squares of the $\tilde{\nu}_R$ and 
$\tilde{\nu}^*_R$,
and of the $\tilde{s}_L$ and $\tilde{s}^*_L$. In particular, for
$m^2_{\tilde{s}}> 0$ and $ m^2_{\tilde{\nu}}<0$ the breaking will occur
along the $D$-flat direction for $|\tilde{\nu}_R| = |\tilde{\nu}^*_R| $ very
large, with the potential ultimately stabilized by loop corrections or higher
dimensional operators. However, since $m^2_{\tilde{s}}> 0, $ $\tilde{s}_L$ and
$
\tilde{s}^*_L$ will acquire (usually different) TeV-scale expectation values
not associated with the flat direction.

For arbitrary expectation values, the \zpr \ mass terms are
\beq
{\cal L} = g'^2 \left( -\frac{5}{2\sqrt{10}} Z_\chi + \frac{1}{\sqrt{24}} Z_\psi \right)^2
 \left( |\tilde{\nu}_R|^2 + |\tilde{\nu}^*_R|^2 \right)
+ g'^2 \left(  \frac{4}{\sqrt{24}} Z_\psi \right)^2
 \left( |\tilde{s}_L|^2 + |\tilde{s}^*_L|^2  \right).
 \eeq
For the breaking pattern described above, this will imply that
$Z_2 \equiv   -\frac{5}{2\sqrt{10}} Z_\chi + \frac{1}{\sqrt{24}} Z_\psi $
will acquire a superheavy mass, while the orthogonal combination
$Z_1 \equiv   \frac{1}{\sqrt{24}} Z_\chi + \frac{5}{2\sqrt{10}} Z_\psi $
will remain at the TeV scale. $Z_1$ decouples from \nur \ and therefore
evades the nucleosynthesis and supernova constraints. One can
even generate a small Dirac \nur \ mass from higher dimensional
operators associated with the $Z_2$ scale\footnote{At least
one $\tilde{\nu}_R$ per family should have a positive mass-squared
to avoid large lepton-Higgsino mixing.}. This
mechanism applies to more general $U(1)' \times U(1)'$ models, indicating
that small Dirac masses are allowed provided certain conditions on
the soft scalar mass squares are satisfied.
 
\section{A Comment on Majorana Phases in Neutrinoless Double Beta Decay}
It is well known that if the neutrinos are Majorana, then the effective
Majorana mass
\beq
M_{ee} = \sum_{i=1}^3 U^2_{ei} m_i,
\label{effmass}
\eeq
where $U$ is the neutrino mixing matrix and $m_i$ is the $i^{th}$
Majorana mass eigenstate,
is observable in neutrinoless double beta decay ($\beta \beta_{0\nu}$).
For Majorana neutrinos, $U$ contains two additional CP-violating
Majorana phases not observable in oscillation experiments, 
in addition to the phase analogous to the one in the quark sector~\cite{numass}.
In principle these can be observable in $\beta \beta_{0\nu}$ if one
knows the masses and mixing angles from other sources.

Unfortunately, it is unlikely that a significant measurement can be made
with forseeable techniques~\cite{phase}. To illustrate this, let us assume
maximal atmospheric mixing and neglect $U_{e3}$ (relaxing these
assumptions does not change the conclusion). Then,
in a basis for which $m_1<m_2<m_3$, one finds
\bea
M_{ee} &=& | c^2 m_1 + s^2 m_2 e^{i\phi}| \,, \ \ \ {\rm~(normal)} \,,
\label{Mee1} \nonumber \\
&=& |c^2 m_2 + s^2 m_3 e^{i\phi}| \,, \ \ \ {\rm~(inverted)},
\label{Mee2}
\eea
for the normal and inverted hierarchies~\cite{patterns}, respectively,
where $c$ and $s$ are the cosine and sine of the Solar mixing angle and
$\phi$ is the Majorana phase.
Proposed $\beta \beta_{0\nu}$ experiments will mainly be sensitive to
the inverted hierarchy or to the normal hierarchy in the case
that the masses are nearly degenerate and large compared to the
splittings. In principle, one could determine $\phi$ from
 sufficiently precise measurements of $M_{ee}$, $m_i$, and the
Solar mixing angle. However, it is shown in~\cite{phase} that
even for extremely optimistic assumptions one will not be able
to distinguish a nontrivial phase from 0 or $ \pi$ at better than $1 \sigma$
(it may be marginally possible to distinguish 0 from $ \pi$).
The biggest problem is that $M_{ee}$ is extracted from a nuclear transition
rate involving a matrix element that must be taken from
theory. Even assuming a breakthrough in reliability
to a 50\% theoretical uncertainty
 (from the
  present factor of 3~\cite{vogel}), the resulting $50$\% error
  on $M_{ee}$ will lead to the discouraging result described above.
  (This can easily be seen, e.g., by taking  a fixed Solar mixing
  in the LMA region and assuming nearly degenerate masses.)

\section*{Acknowledgments}
It is a pleasure to thank the conference organizers for support.
Supported in part by Department of Energy grant DOE-EY-76-02-3071.


\begin{thebibliography}{0}

\bibitem{seesaw}
M. Gell-Mann, P. Ramond, and R. Slansky, in {\it
Supergravity}, ed. F. van Nieuwenhuizen and D. Freedman, (North
Holland, Amsterdam, 1979) p. 315; T. Yanagida, {\it Proc. of the
Workshop on Unified Theory and the Baryon Number of the Universe},
KEK, Japan, 1979; S. Weinberg, 
Phys.\ Rev.\ Lett.\  {\bf 43}, 1566 (1979).
 
\bibitem{leptogenesis}
M.~Fukugita and T.~Yanagida,
Phys.\ Lett.\ B {\bf 174}, 45 (1986).
For a review, see
W.~Buchmuller and M.~Plumacher,
Int.\ J.\ Mod.\ Phys.\ A {\bf 15}, 5047 (2000).


\bibitem{lopsided}
K.~S.~Babu and S.~M.~Barr,
Phys.\ Lett.\ B {\bf 525}, 289 (2002);
I.~Dorsner and S.~M.~Barr,
Nucl.\ Phys.\ B {\bf 617}, 493 (2001).


\bibitem{textures}
For a recent review, see
W. Buchmuller,
hep-ph/0204288.

\bibitem{typeII}
R.~N.~Mohapatra and G.~Senjanovic,
Phys.\ Rev.\ D {\bf 23}, 165 (1981);
C.~Wetterich,
Nucl.\ Phys.\ B {\bf 187}, 343 (1981);
R.~N.~Mohapatra, A.~Perez-Lorenzana and C.~A.~de Sousa Pires,
Phys.\ Lett.\ B {\bf 474}, 355 (2000);
B.~Bajc, G.~Senjanovic and F.~Vissani,
hep-ph/0110310;
B.~Bajc, G.~Senjanovic and F.~Vissani,
Phys.\ Rev.\ Lett.\  {\bf 90}, 051802 (2003);
R.~N.~Mohapatra, M.~K.~Parida and G.~Rajasekaran,
hep-ph/0301234;
H.~S.~Goh, R.~N.~Mohapatra and S.~P.~Ng,
hep-ph/0303055.

\bibitem{sterile}
S.M. Bilenky and B. Pontecorvo,
Lett. Nuovo Cimento {\bf 17}, 569             (1976);
V. Barger, P. Langacker, J. P. Leveille and S. Pakvasa, Phys. Rev. Lett. {\bf 45}, 692 (1980);
S.M.  Bilenky, J. Hosek and S.T. Petcov, Phys. Lett. B {\bf 94}, 495 (1980);
J. Schechter and J.W. F. Valle, Phys. Rev. D {\bf 22}, 2227 (1980);
T.P. Cheng and L.F. Li, Phys. Rev. D {\bf 22}, 2860 (1980); 
T. Yanagida and M. Yoshimura, Progr. Th. Phys. {\bf 64}, 1870 (1980);
I. Yu. Kobzarev et al., Sov. J. Nucl. Phys. {\bf 32}, 823 (1980).   




\bibitem{fournu}
See, for example,
M.~Maltoni, T.~Schwetz, M.~A.~Tortola and J.~W.~Valle,
Nucl.\ Phys.\ Proc.\ Suppl.\  {\bf 114}, 203 (2003);
M.~C.~Gonzalez-Garcia, M.~Maltoni and C.~Pena-Garay,
hep-ph/0108073;
P.~Di Bari,
astro-ph/0302433. 
For a different point of view, see
R.~Foot,
hep-ph/0210393;
H.~Pas, L.~g.~Song and T.~J.~Weiler,
hep-ph/0209373.

\bibitem{LSND}
A.~Aguilar {\it et al.}  [LSND Collaboration],
Phys.\ Rev.\ D {\bf 64}, 112007 (2001).

\bibitem{led1} N.~Arkani-Hamed, S.~Dimopoulos and G.~R.~Dvali,
Phys.\ Lett.\ B {\bf 429}, 263 (1998);
I.~Antoniadis, N.~Arkani-Hamed, S.~Dimopoulos and G.~R.~Dvali,
Phys.\ Lett.\ B {\bf 436}, 257 (1998);
K. R.~Dienes, E.~Dudas and T.~Gherghetta,
Phys.\ Lett.\ B {\bf 436}, 55 (1998).

\bibitem{LED}
See, for example,
K.~R.~Dienes, E.~Dudas and T.~Gherghetta,
Nucl.\ Phys.\ B {\bf 557}, 25 (1999);
G.~R.~Dvali and A.~Y.~Smirnov,
Nucl.\ Phys.\ B {\bf 563}, 63 (1999);
R.~N.~Mohapatra and A.~Perez-Lorenzana,
Nucl.\ Phys.\ B {\bf 576}, 466 (2000),
B {\bf 593}, 451 (2001);
R.~Barbieri, P.~Creminelli and A.~Strumia,
Nucl.\ Phys.\ B {\bf 585}, 28 (2000);
A.~Lukas, P.~Ramond, A.~Romanino and G.~G.~Ross,
Phys.\ Lett.\ B {\bf 495}, 136 (2000);
K.~Agashe and G.~H.~Wu,
Phys.\ Lett.\ B {\bf 498}, 230 (2001);
D.~O.~Caldwell, R.~N.~Mohapatra and S.~J.~Yellin,
Phys.\ Rev.\ D {\bf 64}, 073001 (2001);
C.~S.~Lam,
Phys.\ Rev.\ D {\bf 65}, 053009 (2002);
A.~Ioannisian and A.~Pilaftsis,
Phys.\ Rev.\ D {\bf 62}, 066001 (2000);
N. Arkani-Hamed, S. Dimopoulos, G. R. Dvali and J. March-Russel,
Phys.\ Rev.\ {\bf D65}, 024032 (2002);
T. Appelquist, B. Dobrescu, E. Ponton and H. Yee,
Phys.\ Rev.\ {\bf D65}, 105019 (2002).


\bibitem{DLP}
H.~Davoudiasl, P.~Langacker and M.~Perelstein,
Phys.\ Rev.\ D {\bf 65}, 105015 (2002).

\bibitem{TEVseesaw}
See, for example,
R.~N.~Mohapatra and J.~W.~Valle,
Phys.\ Rev.\ D {\bf 34}, 1642 (1986);
E.~Ma,
Phys.\ Rev.\ D {\bf 66}, 037301 (2002);
and references therein.
For an extension to \upr \ models,
see
A.~E.~Faraggi,
Phys.\ Lett.\ B {\bf 245}, 435 (1990);
J. Kang, P. Langacker, and T. Li, UPR-1010-T, to appear.
For theories with dynamical symmetry breaking, see
T.~Appelquist and R.~Shrock,
Phys.\ Lett.\ B {\bf 548}, 204 (2002) and
hep-ph/0301108.


\bibitem{HDO}
P.~Langacker,
Phys.\ Rev.\ D {\bf 58}, 093017 (1998).

\bibitem{loop} 
D.~Chang and A.~Zee,
Phys.\ Rev.\ D {\bf 61}, 071303 (2000) and references theirin.

\bibitem{numass}
For a recent review, see
 M.~C.~Gonzalez-Garcia and Y.~Nir,
hep-ph/0202058.

\bibitem{BLL}
V.~Barger, P.~Langacker and H.~S.~Lee,
Phys. Rev. {\bf D}, in press, hep-ph/0302066.


\bibitem{phase}
V.~Barger, S.~L.~Glashow, P.~Langacker and D.~Marfatia,
Phys.\ Lett.\ B {\bf 540}, 247 (2002).

\bibitem{gravitons}
For a review, see
J.~Hewett and M.~Spiropulu,
Ann.\ Rev.\ Nucl.\ Part.\ Sci.\  {\bf 52}, 397 (2002);
S. Hannestad and  G. G. Raffelt, hep-ph/0304029.

\bibitem{cavendish}
C.~D.~Hoyle, U.~Schmidt, B.~R.~Heckel, E.~G.~Adelberger, 
J.~H.~Gundlach, D.~J.~Kapner and H.~E.~Swanson,
Phys.\ Rev.\ Lett.\  {\bf 86}, 1418 (2001).

\bibitem{cosmology}
K.~Abazajian, G.~M.~Fuller and M.~Patel,
Phys.\ Rev.\ Lett.\  {\bf 90}, 061301 (2003);
H.~S.~Goh and R.~N.~Mohapatra,
Phys.\ Rev.\ D {\bf 65}, 085018 (2002).

\bibitem{SNbound}
R. Barbieri et al.~\cite{LED}; A. Lukas et al.~\cite{LED}.

\bibitem{ell}
For  recent surveys, see
J.~Erler, P.~Langacker and T.~J.~Li,
Phys.\ Rev.\ D {\bf 66}, 015002 (2002);
S.~Hesselbach, F.~Franke and H.~Fraas,
Eur.\ Phys.\ J.\ {\bf C23}, 149 (2002).



\bibitem{string}
M.~Cveti\v c and P.~Langacker,
Phys.\ Rev.\ D {\bf 54}, 3570 (1996) and
Mod.\ Phys.\ Lett.\ A {\bf 11}, 1247 (1996).

\bibitem{review}
For a review, see, M.~Cveti\v c and P.~Langacker,
in {\it Perspectives on supersymmetry}, ed. G. L. Kane
(World, Singapore, 1998), p. 312.

\bibitem{DSB}
For a review, see
C.~T.~Hill and E.~H.~Simmons,
hep-ph/0203079.

\bibitem{NMSSM}
J.~Ellis, J.~F.~Gunion, H.~E.~Haber, L.~Roszkowski, and
F.~Zwirner, Phys.\ Rev.\ D {\bf 39}, 844 (1989); and references
therein. Recent references may be found in U. Ellwanger,
J. F. Gunion, and C. Hugonie, hep-ph/0111179.


\bibitem{muprob}
D.~Suematsu and Y.~Yamagishi,
Int.\ J.\ Mod.\ Phys.\ A {\bf 10}, 4521 (1995);
M.~Cveti\v c, D.~A.~Demir, J.~R.~Espinosa, L.~L.~Everett and P.~Langacker,
Phys.\ Rev.\ D {\bf 56}, 2861 (1997)
[Erratum-ibid.\ D {\bf 58}, 119905 (1997)].


\bibitem{general}
J.~Erler,
Nucl.\ Phys.\ B {\bf 586}, 73 (2000).

 \bibitem {gaugeupr}
P.~Langacker, N.~Polonsky and J.~Wang,
Phys.\ Rev.\ D {\bf 60}, 115005 (1999).

\bibitem{explim}  F. Abe { et al.} [CDF Collaboration],
{ Phys. Rev.  Lett.} {\bf 79}, 2192 (1997).

\bibitem{LEPmixing}
R.~Barate {\it et al.}  [ALEPH Collaboration],
Eur.\ Phys.\ J.\ C {\bf 12}, 183 (2000);
P.~Abreu {\it et al.}  [DELPHI Collaboration],
Phys.\ Lett.\ B {\bf 485}, 45 (2000).

\bibitem{indirect}
J.~Erler and P.~Langacker,
Phys.\ Lett.\ B {\bf 456}, 68 (1999), 
Phys.\ Rev.\ Lett.\  {\bf 84}, 212 (2000), and references therein.


\bibitem{NuTeV}
G.~P.~Zeller { et al.}  [NuTeV Collaboration],
Phys.\ Rev.\ Lett.\  {\bf 88}, 091802 (2002).

\bibitem{nonuniversal} 
P.~Langacker and M.~Plumacher,
Phys.\ Rev.\ D {\bf 62}, 013006 (2000) and references theirin.

\bibitem{afb}
The LEP Electroweak Working Group and SLD Heavy Flavour Group,
hep-ex/0212036.

\bibitem{APV} C.S. Wood { et al.}, { Science} {\bf 275}, 1759
(1997);
S.C. Bennett and C.E. Wieman, { Phys. Rev. Lett.} {\bf 82}, 2484
(1999). 

\bibitem{interp}
V.A. Dzuba, V.V. Flambaum, and J.S.M. Ginges,
Phys.\ Rev.\ D {\bf 66}, 076013 (2002);
M. Y. Kuchiev,
J.\ Phys.\ B {\bf 35}, L503 (2002).



\bibitem{collider}
For reviews, see
M.~Cveti\v c and S.~Godfrey,
hep-ph/9504216;
A.~Leike,
Phys.\ Rept.\  {\bf 317}, 143 (1999).
For a recent update, see S.~Godfrey
in {\it Proc. of the APS/DPF/DPB Summer Study
on the Future of Particle Physics (Snowmass 2001) } ed. N.~Graf,
hep-ph/0201093 and hep-ph/0201092.

\bibitem{LW} See, for example,
P.~Langacker and J.~Wang,
Phys.\ Rev.\ D {\bf 58}, 115010 (1998).

\bibitem{CCEEL} 
G.~Cleaver, M.~Cvetic, J.~R.~Espinosa, L.~L.~Everett, P.~Langacker and J.~Wang,
Phys.\ Rev.\ D {\bf 59}, 055005 (1999); {\bf 59}, 115003 (1999), and
references theirin.


\bibitem{CSU}
M.~Cvetic, G.~Shiu and A.~M.~Uranga,
Phys.\ Rev.\ Lett.\  {\bf 87}, 201801 (2001);
Nucl.\ Phys.\ B {\bf 615}, 3 (2001).

\bibitem{CLS}
M.~Cvetic, P.~Langacker and G.~Shiu,
Phys.\ Rev.\ D {\bf 66}, 066004 (2002).

\bibitem{flatness}
G.~Cleaver, M.~Cvetic, J.~R.~Espinosa, L.~L.~Everett and P.~Langacker,
Nucl.\ Phys.\ B {\bf 525}, 3 (1998); {\bf 545}, 47 (1999).


\bibitem{secluded} See the first article in~\cite{ell}.

\bibitem{intermediate}
G.~Cleaver, M.~Cvetic, J.~R.~Espinosa, L.~L.~Everett and P.~Langacker,
Phys.\ Rev.\ D {\bf 57}, 2701 (1998).

\bibitem{susylowmass}
N.~Arkani-Hamed, L.~J.~Hall, H.~Murayama, D.~R.~Smith and 
         N.~Weiner,  Phys. Rev. D64  115011 (2001);
F. Borzumati and Y. Nomura, Phys. Rev. D64, 053005 (2001); 
F.~Borzumati, K.~Hamaguchi, Y.~Nomura and T.~Yanagida,
hep-ph/0012118.


\bibitem{triplet}
T.~Hambye, E.~Ma and U.~Sarkar,
Nucl.\ Phys.\ B {\bf 602}, 23 (2001);
E.~Ma and U.~Sarkar,
Phys.\ Rev.\ Lett.\  {\bf 80}, 5716 (1998).

\bibitem{smallDirac}
P.~Q.~Hung,
Phys.\ Rev.\ D {\bf 62}, 053015 (2000);
R.~Kitano,
Phys.\ Lett.\ B {\bf 539}, 102 (2002).



\bibitem{steigman}
G.~Steigman, K.~A.~Olive and D.~N.~Schramm,
Phys.\ Rev.\ Lett. {\bf 43}, 239 (1979);
K.~A.~Olive and D.~N.~Schramm and G.~Steigman,
Nucl.\ Phys.\ {\bf B180}, 497 (1981).

\bibitem{dolgov}
For a general review of neutrinos in cosmology, see
A.~D.~Dolgov,
Phys.\ Rept.\ {\bf 370}, 333 (2002).


\bibitem{supernova}
G. Raffelt and D. Seckel,
Phys.\ Rev.\ Lett. {\bf 60}, 1793 (1988);
R. Barbieri and R. N. Mohapatra,
Phys.\ Rev.\ {\bf D39}, 1229 (1989);
J. A. Grifols and E. Masso,
Nucl.\ Phys.\ {\bf B331}, 244 (1990);
J.~A.~Grifols, E.~Masso and T.~G.~Rizzo,
Phys.\ Rev.\ D {\bf 42}, 3293 (1990);
T.~G.~Rizzo,
Phys.\ Rev.\ D {\bf 44}, 202 (1991).


\bibitem{decouplingmodel}
P. Langacker, in preparation.


\bibitem{patterns}
Y.~Farzan, O.~L.~Peres and A.~Y.~Smirnov,
Nucl.\ Phys.\ B {\bf 612}, 59 (2001);
W.~Rodejohann, hep-ph/0203214;
S.~Pascoli and S.~T.~Petcov,
Phys.\ Lett.\ B {\bf 544}, 239 (2002).

\bibitem{vogel}
S.~R.~Elliott and P.~Vogel,
Ann.\ Rev.\ Nucl.\ Part.\ Sci.\  {\bf 52}, 115 (2002).


\end{thebibliography}
\end{document}